%% file: paper.tex
\documentclass[runningheads]{llncs}        
\usepackage{amssymb}
\usepackage{amsmath}

\usepackage{amsthm}
\usepackage{bussproofs}
\usepackage{stmaryrd}

\usepackage{dsfont}

\usepackage{bbm}
\usepackage{graphicx}
\usepackage{listings}
\usepackage{bussproofs}
\usepackage{tikz}
\usepackage{floatflt} 

\usepackage{ifthen}
\usepackage{xspace}
\usepackage{pdfpages}
\usepackage{todonotes}
\usepackage[inline]{enumitem}

\usepackage{floatflt} 
\usepackage[hidelinks]{hyperref}

\usepackage{pgf-umlsd}
\usetikzlibrary{arrows,shapes,trees,positioning,decorations,shapes}
\usetikzlibrary{decorations.markings,automata}

\input{abs}

\input{definitions}

\title{\texttt{Crowbar}: Behavioral~Symbolic~Execution for Deductive~Verification of Active~Objects}
\author{Eduard Kamburjan\orcidID{0000-0002-0996-2543}\inst{1} \and Marco Scaletta\orcidID{0000-0001-5298-3369}\inst{2} \and Nils Rollshausen\orcidID{0000-0003-2445-8684}\inst{2}}
\titlerunning{\texttt{Crowbar}: Behavioral Symbolic Execution for Active Objects}

\institute{
University of Oslo, Oslo, Norway\\
 \email{eduard@ifi.uio.no}
\and
Technische Universit{\"a}t Darmstadt, Darmstadt, Germany\\
 \email{scaletta@cs.tu-darmstadt.de, nils.rollshausen@stud.tu-darmstadt.de}
}

\begin{document}
\maketitle

\begin{abstract}
We present the \crowbar tool, a deductive verification system for the \ABS language.
\ABS models distributed systems with the Active Object concurrency model.
\crowbar implements \emph{behavioral} symbolic execution: each method is symbolically executed, 
but specification and prior static analyses influence the shape of the symbolic execution tree.
User interaction is realized through guided counterexamples, which present failed proof branches in terms of the input program.
\crowbar has a clear interface to implement new specification languages and verification calculi in the Behavioral Program Logic
and has been applied for the biggest verification case study of Active Objects.
\end{abstract}

\section{Introduction}\label{sec:intro}
\input{intro}

\section{Preliminaries: \ABS and \BPL}\label{sec:prelim}
\input{prelim}
\section{Behavioral Symbolic Execution and Structure}\label{sec:structure}
\input{structure}
\section{Front-end: Specification and Usage}\label{sec:frontend}
\input{frontend}
\section{Middle-end: Behavioral Symbolic Execution}\label{sec:middleend}
\input{middleend}

\section{Back-end: Verification}\label{sec:backend}
\input{backend}

\section{Application, Conclusion and Future Work}\label{sec:conclusion}
\input{conclusion}

\subsubsection*{Acknowledgments}
This work was partially funded by the Hessian LOEWE initiative within the Software-Factory 4.0 project.
We thank Daniel Drodt for his work on nullability checks. 

\bibliographystyle{acm}
\bibliography{ref}

\end{document}

%% file: abs.tex
\DeclareOldFontCommand{\rm}{\normalfont\rmfamily}{\mathrm}
\DeclareOldFontCommand{\sf}{\normalfont\sffamily}{\mathsf}
\DeclareOldFontCommand{\tt}{\normalfont\ttfamily}{\mathtt}
\DeclareOldFontCommand{\bf}{\normalfont\bfseries}{\mathbf}
\DeclareOldFontCommand{\it}{\normalfont\itshape}{\mathit}
\DeclareOldFontCommand{\sl}{\normalfont\slshape}{\@nomath\sl}
\DeclareOldFontCommand{\sc}{\normalfont\scshape}{\@nomath\sc}
\DeclareRobustCommand*\cal{\@fontswitch\relax\mathcal}
\DeclareRobustCommand*\mit{\@fontswitch\relax\mathnormal}

\colorlet{keywordcolor}{blue!50!black}
\colorlet{commentcolor}{green!60!black}
\colorlet{typecolor}{violet}
\newcommand{\sourcefont}{\ttfamily\small}
\newcommand{\commentfont}{\slshape\rmfamily\color{commentcolor}}
\lstdefinelanguage{ABS}{
        keywords={od,do,assert,this,new,type,def,case,of,local,class,interface,
        extends,implements,if,then,else,await,get,Fut,return,skip,while,module,
        import,export,from,to,suspend,delta,adds,modifies,removes,original,productline,
        features,core,corefeatures,optionalfeatures,after,when,product,hasAttribute,
        hasMethod,hasField,hasInterface,uses,root,extension,group,allof,oneof,require,
        stateupdate,object,main,objectupdate,classupdate,fi,
        exclude,original,ifin,ifout,opt,null,%critical,port,rebind,duration,deadline,now,
        newgroup,thiscomp,in,joins,leaves,subtypeOf,acquire,except,as,component,Pre,Abs
        },
        keywordstyle=\color{keywordcolor}\bfseries\sffamily,
        morekeywords=[2]{Unit, Int, Bool, Rat, List, Set, Pair, Fut, Maybe, String, Triple, Either, Map},
        keywordstyle=[2]\color{typecolor},
        sensitive=true,
        comment=[l]{//},
        morecomment=[s]{/*}{*/},
        morestring=[b]"
}

\lstdefinelanguage[v9]{Java}[]{Java}{
        morekeywords={module,requires,provides,uses,with,to,exports}
}
\lstdefinelanguage[ContextJ]{Java}[]{Java}{
        morekeywords={layer,with,without,proceed,before,after}
}
\lstdefinelanguage[FOP]{Java}[]{Java}{
        morekeywords={refines,original,Super}
}
\lstdefinelanguage[JastAdd]{Java}[]{Java}{
        morekeywords={aspect,syn,inh,lazy}
}

\lstdefinestyle{code}{
        basicstyle=\sourcefont\upshape,
        keywordstyle=\color{keywordcolor}\bfseries\sffamily,
        commentstyle=\commentfont,
        columns=fullflexible,
        mathescape=true,
        escapechar={\#},
        keepspaces=true,
        showstringspaces=false,
        aboveskip=8pt, % default is \medskipamount
        numbers=left,
        stepnumber=1, 
        numberstyle=\ttfamily\scriptsize\color{gray},
        numbersep=4pt,
        xleftmargin=1.5em,
        xrightmargin=1.5em,
        framexleftmargin=1.2em,
        framexrightmargin=1em,
        framextopmargin=0.5ex,
        breaklines=true,
        breakindent=3pt,
}

\lstdefinestyle{abs}{
        style=code,
        language=ABS,
}
\lstdefinestyle{java}{
        style=code,
            language=Java
}
\lstdefinestyle{java9}{
        style=code,
            language=[v9]Java
}
\lstdefinestyle{aspectj}{
        style=code,
        language=[AspectJ]Java
}
\lstdefinestyle{jastadd}{
        style=code,
        language=[JastAdd]Java
}
\lstdefinestyle{contextj}{
        style=code,
        language=[ContextJ]Java
}
\lstdefinestyle{FOP}{
        style=code,
        language=[FOP]Java
}
\lstdefinestyle{scala}{
        style=code,
        language=Scala,
        morekeywords={self}
}

\newcommand{\code}[2][]{\lstinline[style=code,basicstyle=\ttfamily\upshape,#1]{#2}}

\newcommand{\abs}[2][]{\code[style=abs,#1]{#2}}

\lstnewenvironment{srccode}[1][]{
        \minipage{\linewidth}
        \lstset{style=code,
        backgroundcolor=\color{white},
        rulecolor=\color{gray!50},
        frame=tblr,
        captionpos=b,
        #1}
}{\endminipage}

\lstnewenvironment{abscode}[1][]{
        \minipage{1\linewidth}
        \lstset{style=abs,
        backgroundcolor=\color{white},
        rulecolor=\color{gray!50},
        frame=tblr,
        captionpos=b,
        #1}
}{\endminipage}

\lstnewenvironment{javacode}[1][]{
        \minipage{\linewidth}
        \lstset{style=java,
        backgroundcolor=\color{gray!5},
        rulecolor=\color{gray!50},
        frame=tblr,
        captionpos=b,
        #1}
}{\endminipage}

\lstnewenvironment{jastaddcode}[1][]{
        \minipage{\linewidth}
        \lstset{style=jastadd,
        backgroundcolor=\color{gray!5},
        rulecolor=\color{gray!50},
        frame=tblr,
        captionpos=b,
        #1}
}{\endminipage}

%% file: definitions.tex
\newcommand{\COMMENT}[1]{}

\makeatletter
\newsavebox{\@brx}
\newcommand{\llangle}[1][]{\savebox{\@brx}{\(\m@th{#1\langle}\)}%
  \mathopen{\copy\@brx\kern-0.5\wd\@brx\usebox{\@brx}}}
  \newcommand{\rrangle}[1][]{\savebox{\@brx}{\(\m@th{#1\rangle}\)}%
    \mathclose{\copy\@brx\kern-0.5\wd\@brx\usebox{\@brx}}}
    \makeatother

\let\temp\phi
\let\phi\varphi
\let\varphi\temp

\makeatletter
\newcommand{\xRightarrow}[2][]{\ext@arrow 0359\Rightarrowfill@{#1}{#2}}

\newcommand{\xabs}[1]{\text{\abs{#1}}}

\newcommand{\sem}[1]{\ensuremath{ \llbracket #1 \rrbracket \xspace}}

\newcommand{\sep}{  \ | \ }

\newcommand{\methodname}{\ensuremath{\synt{m}}\xspace}

\COMMENT{

}

\newcommand{\synt}[1]{\ensuremath{\xabs{#1}}}

\newcommand{\statement}{\ensuremath{\xabs{s}}\xspace}

\newcommand{\trace}{\ensuremath{\theta}\xspace}

\newcommand{\crowbar}{\texttt{Crowbar}\xspace}
\newcommand{\ABS}{\texttt{ABS}\xspace}
\newcommand{\Key}{\texttt{KeY}\xspace}
\newcommand{\BPL}{\texttt{BPL}\xspace}

%% file: intro.tex
Symbolic Execution (SE)~\cite{King76,YangFBCW19} is a core technique in program verification, 
which is recently receiving increasing attention for concurrent systems~\cite{SantosMAG20,BoerB19,BoerBJPTT20}.
For the Active Objects~\cite{boer} concurrency model, experiences with multiple case studies~\cite{DinTHJ15,KamburjanH17,wao} led to the design of the
Behavioral Program Logic (\BPL)~\cite{Kamburjan19}, a dynamic logic that generalizes SE to \emph{Behavioral Symbolic Execution} (BSE).

In this paper, we introduce \crowbar, a system for the deductive verification of Active Objects~\cite{boer}, which targets the \ABS language~\cite{abs},
and implements \BPL. %, an modular dynamic logic, whose calculus employs \emph{guided} symbolic execution.
Contrary to other heavyweight%
\footnote{I.e., static logic-based functional verification using loop invariants for unbounded behavior. For a discussion on lightweight vs.\ heavyweight SE we refer to~\cite[Ch.~3]{dominic}.}
symbolic execution approaches, behavioral symbolic execution uses the specification already when building the symbolic execution tree:
the specification is reduced on-the-fly and \crowbar can fails early if the remaining specification cannot be adhered to.
Not only the specification is used to influence the shape of the SE tree, but also static analyses. %, such as nullability checkers.

Additionally, \crowbar exploits the Active Object model to drastically simplify its handling of concurrency:
Active Objects carefully distinguishes between local execution, where a single method essentially behaves as in a sequential system, 
and global concurrency. \crowbar performs its analysis on the local layer, but has a clear interface to external analyses
if a global property is required.

BSE improves automatization of deductive verification, as they reduce the size of the final logical formulas. 
This allows us to scale heavyweight SE for distributed systems:
\crowbar is used to perform the to-date biggest case study~\cite{crowpaper} in deductive verification of Active Objects.

We introduce \emph{guided counterexamples} for user feedback. From a failed proof branch, a \emph{program} is generated that allows the user
to examine the failed proof in terms of the input language and retains the original structure of the method.

\crowbar is implemented with flexibility in mind: one of its long term goals is to explore
the design space of behavioral symbolic execution and specification of distributed systems. \emph{Two} SE calculi are already implemented.
\Key-\ABS~\cite{DinBH15}, the predecessor system for \ABS is superseded by \crowbar and discussed in Sec.~\ref{sec:conclusion}.

We introduce \ABS and \BPL in Sec.~\ref{sec:prelim}, and the structure of \crowbar in Sec.~\ref{sec:structure}, 
before describing front-end (Sec.~\ref{sec:frontend}),  middle-end\footnote{We borrow this term from compiler construction.} 
(Sec.~\ref{sec:middleend}), and back-end (Sec.~\ref{sec:backend}) in detail.
\crowbar is available at \url{https://github.com/Edkamb/crowbar-tool}.

%% file: prelim.tex
Active Objects combine object-orientation with cooperative scheduling and actor-based communication with futures.
For a formal introduction we refer to~\cite{boer,abs}. % and for the symbolic execution rules in \BPL to~\cite{Kamburjan19,sao}.
\begin{description}
\item[Strong Encapsulation.]
Every object is strongly encapsulated at runtime, such that no other object can access its fields, not even objects of the same class.
\item[Asynchronous Calls with Futures.] 
Each method call to another object is asynchronous and generates a future. Futures can be passed around and are used to synchronize on the process generated by the call.
Once the called process terminates, its future is \emph{resolved} and the return value can be retrieved. We say that the process \emph{computes} its future.
\item[Cooperative Scheduling.] 
At every point in time, at most one process is active per object and a running process cannot be interrupted unless it \emph{explicitly} releases the object.
This is done either by termination with a \abs{return} statement or with an \abs{await g} statement that waits until guard \abs{g} holds.
A guard polls whether a future is resolved or whether a boolean condition holds.
\end{description}
Concurrent systems are challenging for SE, but Active Objects allow to perform \emph{local} SE on single methods through their strong encapsulation and decoupling of caller and callee processes. Special care, however, has to be taken to keep track of futures and correct handling of state when using \abs{await}.
We introduce the \ABS language using an example to demonstrate these features. 

\begin{example}
Consider Lst.~\ref{fig:example}.
The class \abs{Monitor} has two fields: \abs{s} which is a server that is monitored and specified as being non-null (more on specification in Sec.~\ref{sec:frontend}) 
and \abs{beats}, a counter for successful requests to \abs{s}.
The method \abs{heartbeat} sends a request to itself (l.~\ref{line1}) by an asynchronous method call \mbox{(by using \abs{!})}.
Afterwards, the return value of the call is retrieved (l.~\ref{line2}, using \abs{get}). This blocks the process until the \abs{httpRequest} process has terminated.
No other process can run on this object until this happens. If the request was successful, \abs{beats} is increased by 1.
Method \abs{reset} waits \emph{without blocking} until the number of success reaches a passed threshold and resets \abs{beats}.
Synchronous calls are possible (l.~\ref{line3}) on \abs{this}. 
\begin{figure}[tb]
\centering
\renewcommand{\figurename}{Listing}
\begin{abscode}
[Spec:Requires(this.s != null)][Spec:ObjInv(this.s != null)]
class Monitor(Server s) {
 Int beats = 0;
 [Spec:Ensures(this.beats >= old(this.beats) && result == this.beats)]
 Int heartbeat() {
  Fut<Int> req = s!httpRequest(); $\label{line1}$
  Int status = req.get; $\label{line2}$
  if(status == 200) { this.beats = this.beats + 1;} 
  else { this.handleError(); } $\label{line3}$
  return this.beats;
 }
 Unit reset(Int i){ await this.beats == i; this.beats = 0; } $\label{line4}$
 Unit handleError() { this.beats = 0; /* ... */ }
}
\end{abscode}
\caption{An example \ABS program with specification.} % \ektodoin{adjust}.} %showcasing counterexample generation.}
%Note that the \emph{handleError} method can modify the heap, thus potentially violating the first \emph{ensures} clause.}
\label{fig:example}
\end{figure}
\end{example}

\paragraph{Verification using Symbolic Execution.}
Symbolic execution describes the execution of a program (or statement) with \emph{symbolic values}. 
A symbolic value is a placeholder and can be described by condition collecting during the symbolic execution.
E.g., at points the control flow branches, knowledge about the guard of the \abs{if} statement describes the involved symbolic value.

Symbolic execution is used as a \emph{proof strategy} for a sequent calculus of first-order dynamic logics:
A sequent of the form $\Gamma \Rightarrow \{U\}[\statement]\phi, \Delta$ 
represents a symbolic state, where \statement is the statement left to symbolically execute, $U$ the state update (i.e., a syntactic representation of accumulated substitusions~\cite{Beckert00}), $\phi$ the postcondition, and $\bigwedge\Gamma \wedge \neg\bigvee\Delta$
describes the accumulated knowledge and path condition. A rule can have, e.g., the following form (SE rules are read bottom-up)
\begin{prooftree}
\AxiomC{$\Gamma, \{U\}\xabs{e} \Rightarrow \{U\}[\xabs{s}_1]\phi, \Delta$}
\AxiomC{$\Gamma, \{U\}\neg\xabs{e} \Rightarrow \{U\}[\xabs{s}_2]\phi, \Delta$}
\BinaryInfC{$\Gamma \Rightarrow \{U\}[\xabs{if(e) s}_1\xabs{else s}_2]\phi, \Delta$}
\end{prooftree}
Heavyweight SE with dynamic logics is successfully applied to discover highly involved bugs in non-concurrent libraries of mainstream languages~\cite{GouwRBBH15}.

\subsection{Behavioral Program Logic.}
The behavioral program logic (BPL)~\cite{Kamburjan19} is a generalization of dynamic logic. % that introduces guided symbolic execution (BSE).
It uses \emph{behavioral} modalities, which we informally introduce now. %, and embeds them into a first-order logic.
One of the shortcomings of symbolic execution with dynamic logics is that they first fully symbolically execute the program and then evaluate the postcondition.
For distributed systems the specification, however, often contains a temporal element and can be partially checked already \emph{during} symbolic execution.

\paragraph{Behavioral Modalities.}
A behavioral (box-)modality uses a behavioral specification $\left(\alpha_{\mathbb{T}},\tau_{\mathbb{T}}\right)$, where $\alpha_{\mathbb{T}}$ maps elements of 
$\tau_{\mathbb{T}}$ to trace formulas
%\footnote{In the original formalization monadic second-order logic is used, but the specific logic is not fixed, as long as its models are program traces.} 
and has the form $\left[\statement \Vdash^{\alpha_{\mathbb{T}}} \tau\right]$ with $\tau\in \tau_{\mathbb{T}}$.
Its semantics expresses partial correctness: a state $\sigma$ satisfies the modality, if every trace of a normally terminating run of $\statement$ from $\sigma$ (the set of these traces is denoted $\sem{\statement}^\sigma$) is a model for the trace formula $\alpha_{\mathbb{T}}(\tau)$:
\[\sigma \models \left[\statement \Vdash^{\alpha_{\mathbb{T}}} \tau\right] \iff \forall \trace \in\sem{\statement}^\sigma.~ \trace \models \alpha_{\mathbb{T}}(\tau)\]
E.g., for the behavioral specification of post-conditions $\mathbb{T}_{\mathsf{pst}} = \left(\alpha_{\mathsf{pst}},\tau_{\mathsf{pst}}\right)$, we use all \BPL formulas as elements of the specification language $\tau_{\mathsf{pst}}$. The semantics extracts the last state for a trace.
\[\tau_{\mathsf{pst}} = \{\phi \sep \phi \text{ is a \BPL Formula}\} \qquad\qquad \trace \models \alpha_{\mathsf{pst}}(\phi) \iff \mathsf{last}(\trace) \models \phi\]

Behavioral specifications separate \emph{syntax} of the specification ($\tau$) and its semantics as a trace specification ($\alpha$).
This distinction enables behavioral symbolic execution: the \emph{design} of $\tau$ 
can now aim to have a simple proof calculus that is not restricted by the structure of the logic underlying $\alpha$. Furthermore, $\tau$ can serve as an interface to external analyses. 

Several behavioral specification frameworks are formalized~\cite{thesisek}, based on logics (post-conditions, LTL), behavioral types (effect types, session types~\cite{HondaYC08}), static analyses (points-to) and specification techniques (cooperative method contracts, object invariants). The calculi for these specifications can refer to each other: the post-condition calculus is used for loop invariants in other calculi.

\paragraph{Symbolic Execution in \BPL.}
%We now make precise what we mean by designing a representation for guided symbolic execution.
%Heavyweight symbolic execution is formalized as a sequent calculus for dynamic logic.
As mentioned above, established calculi for dynamic logics (like JavaDL~\cite{keybook}, ABSDL~\cite{DinO15} or DTL~\cite{BeckertB13}) perform symbolic execution
by reducing the statement inside a modality \emph{without considering the post-condition}. 
In contrast, for each step in a behavioral type system~\cite{beh}, the statement is matched with the current specification.
\BPL combines both: a logical framework with a behavioral type system-style calculus.

\begin{example}\label{ex:session}
We give a simple example using our session type system~\cite{KamburjanC18}: the following expresses that the statement \statement first calls method \abs{m} on role \abs{f} and then \abs{n}
on role \abs{g}. Finally, the last action is a \abs{return} in a state where \abs{result == 0} holds. No other communication, synchronization or suspension happens.
\[\big[ \statement \Vdash^{\alpha_{\mathsf{ST}}} \mathbf{f}\xabs{!m.}\mathbf{g}\xabs{!n}.\downarrow \xabs{result == 0}\big]\]
One (slightly prettified) rule for session types is the following. 
\begin{prooftree}
\AxiomC{$\Gamma \Rightarrow \{U\}\xabs{this.f} \doteq \xabs{r}, \Delta$}
\AxiomC{$\Gamma \Rightarrow \{U\}\big[\xabs{s} \Vdash^{\alpha_{\mathsf{ST}}} \xabs{L}\big], \Delta$}
\BinaryInfC{$\Gamma \Rightarrow \{U\}\big[\xabs{this.f!m()}\xabs{;s} \Vdash^{\alpha_{\mathsf{ST}}} \xabs{r!m.L}\big], \Delta$}
\end{prooftree}
The first premise checks that the role and field coincide, the second premise \emph{reduces the type} during the symbolic execution step.
Note that this premise contains no modality -- we call such branches \emph{side-branches}.
We stress that the conclusion of the rule has to match (1) the call in specification and statement and (2) the method names both \emph{syntactically}.
If matching fails, SE stops.
\end{example}

%In~\cite{thesisek}, it is discussed that a session type can be represented as a logical trace formula, but that this representation is not suited for automatization.
%A proper representation $\tau_\mathbb{T}$ can simplify drastically the symbolic execution calculus and allows one to include guides.
\crowbar implements 2 calculi: (1) cooperative contracts~\cite{sao} with object invariants and (2) a session type system for Active Objects~\cite{KamburjanC18}.

%% file: structure.tex
\begin{figure}[t]
\begin{minipage}{0.375\textwidth}
\begin{abscode}
class C([NonNull]I f){
  I g;
  [Spec: $\mathbf{f}\mathtt{!m.}\mathbf{g}\mathtt{!n}.\downarrow \phi$]
  Unit m(){
    this.f!m(); 
    this.g!m(); 
    this.f!m(); 
    return 0;
  }
}
\end{abscode}
\end{minipage}
\begin{minipage}{0.5\textwidth}
\begin{tikzpicture}
\node[very thick, draw=black,rectangle,fill=none] at (0,-0.5) (A) {
    \begin{minipage}{4cm}
    \abs{this.f!m(); this.g!m(); this.f!m(); return 0;}
    \end{minipage}
    $\Vdash^{\alpha_{\mathsf{ST}}} \mathbf{f}\xabs{!m().}\mathbf{g}\xabs{!n}.\downarrow \phi$
    };
\node[very thick, draw=black,rectangle,fill=none] at (0,1) (B) {$ \xabs{this.g!m(); this.f!m(); return 0;} \Vdash^{\alpha_{\mathsf{ST}}} \mathbf{g}\xabs{!n}.\downarrow \phi$};
\node[very thick, draw=black,rectangle,fill=none] at (-1.0,1.75) (C) {$ \xabs{this.f!m(); return 0;} \Vdash^{\alpha_{\mathsf{ST}}} \downarrow \phi$};
\node[very thick, dashed,draw=black,rectangle,fill=none] at (-1.0,2.5) (F) {$ \xabs{return 0;} \Vdash^{\alpha_{\mathsf{ST}}} \lightning$};
\node[dashed, draw=black,rectangle,fill=none] at (-1.0,3.25) (G) {$\phi \lightning$};
\node[dashed, draw=black,rectangle,fill=none] at (2.4,0.35) (D) {{\tiny\bf (1)} \xabs{this.f != null}};
\node[draw=black,rectangle,fill=none] at (2.6,1.75) (E) {\xabs{this.g != null}};
\node[dashed, draw=black,rectangle,fill=none] at (2.6,2.5) (H) {\xabs{this.f != null}};
\draw (A) edge[dashed] (D);
\draw (B) edge (E);
\draw (C) edge[dashed] (F);
\draw (C) edge[dashed] (H);
\draw (A) edge (B);
\draw (B) edge (C);
\draw (C) edge[dashed] (F);
\draw (F) edge[dashed] (G);
\end{tikzpicture}
\end{minipage}
\caption{Illustration of guides in BSE. Verification of the method to the left results in a symbolic execution tree where the dashed nodes are not generated.}
\label{fig:gsetree}
\end{figure}
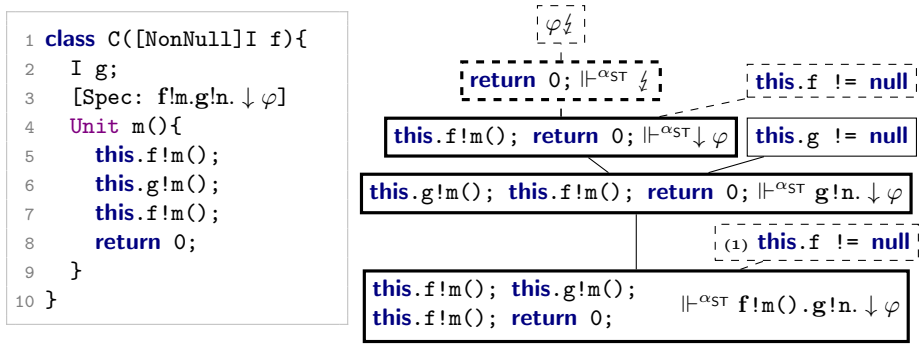

\paragraph{Behavioral Symbolic Execution.}
Before describing BSE in more detail in Sec.~\ref{sec:middleend}, we use the short example in Fig.~\ref{fig:gsetree} to illustrate the difference between SE and BSE: 
the node marked with $(1)$ is not generated, as the type system ensures that \abs{this.f} is always non-\abs{null} (due to being annotated with \abs{NonNull}).
The other dashed nodes are omitted because the method is specified to make two calls, but executes three. 
It does not check any steps after the second call as these are already following a wrong execution path.
We omitted (a) the update, (b) the collected path condition and (c) the role check branches in the illustration.

\begin{figure*}[bt]
\includegraphics[scale=0.35]{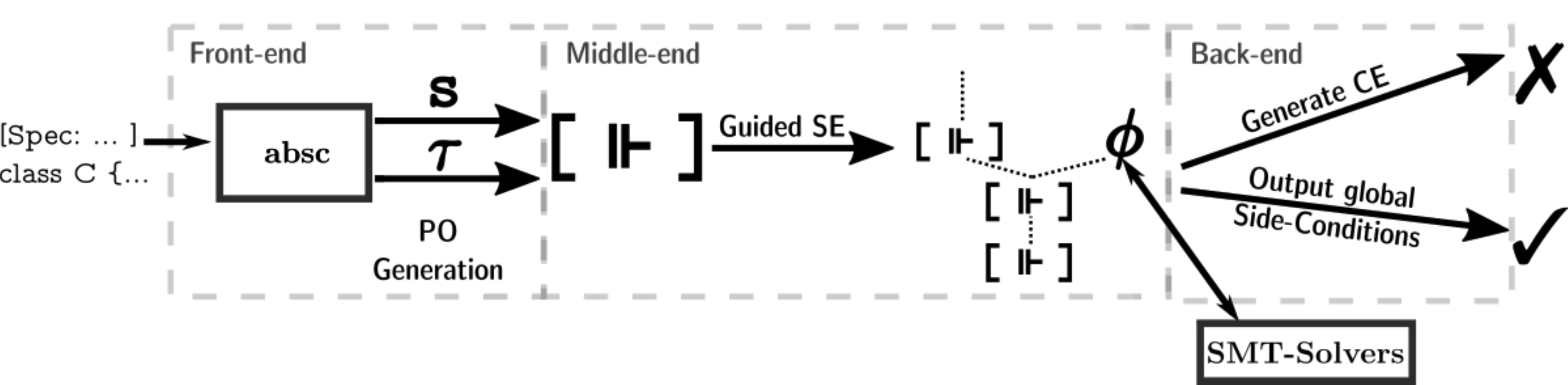}\vspace{-3mm}
\caption{Structure of \crowbar. The \ABS compiler \texttt{absc} and the SMT-solvers are external.}
\label{fig:structure}
\end{figure*}

\paragraph{Structure.}
\crowbar has the structure illustrated in Fig.~\ref{fig:structure} and can be divided into front-end, middle-end
and back-end.
We describe each component on a high level here and give details in the next sections.

The front-end contains the \ABS parser and type-checker, that takes an \ABS file annotated with specification, and outputs for each method (1) its method body \statement and (2) a suitable specification $\tau$. %The information from lightweight analyses is part of $\tau$.
The statement \statement is transformed into the internal representation of \crowbar and the generated proof obligation (PO) is a sequent of the form $\phi \Rightarrow \{U\}[\statement \Vdash^\alpha \tau]$ for fitting $\alpha$, $U$ and $\phi$.

The middle-end performs behavioral symbolic execution and outputs a symbolic execution tree where the leaves are symbolic nodes (where symbolic execution could not continue), logic nodes (where a modality-free formula has to be proven valid) or static nodes (where external analyses must be performed).
This part is modular to enable the implementation of further BSE approaches. 

The back-end passes logic nodes to SMT solvers in the SMT-LIB format 
%\footnote{This deeper integration of SE and SMT, which we discuss in Sec.~\ref{sec:backend}, is one of the main conclusions from the \Key-Java case study from verifying OpenJDK's TimSort~\cite{GouwBBHRS19}.}  
and outputs static nodes to the user, who can pass them to their respective analysis. 
Symbolic nodes as leaves in the final tree always fail the verification.
Additionally, the back-end transforms the symbolic execution tree of a failed proof into guided counterexamples in \ABS
and enhaces it with the values retrieved from the model of the SMT solver.
By using guided counterexamples, the user is not exposed to the internals of \crowbar and user-interactions are all on the level of the \ABS.

%% file: frontend.tex
\crowbar specifications are part of the input \ABS file using \emph{\abs{Spec}-annotations}. 
A method can be annotated with \abs{Ensures} and \abs{Requires} as post and preconditions.
In interfaces, these specifications can only contain parameters (and \abs{result}, the special variable for the return value).
Objects are annotated with creation conditions (\abs{Requires}) and invariants (\abs{ObjInv}).
Additionally, loops are annotated with loop invariants using \abs{WhileInv}.
For an example, we refer to Fig.~\ref{fig:example}.

Following the principles behind the  Java Modeling Language (JML)~\cite{jml}, \crowbar supports \abs{old} to refer to the pre-state of the method and \abs{last} to refer to the prestate of the last suspension. 
Additionally, \crowbar supports the \abs{Succeeds} and \abs{Overlaps} context sets~\cite{sao}: if the precondition of a method contains assertions about the heap, then it is not clear which method is responsible to establish it --- due to the concurrency model, the caller has no control over the fields of the callee object. 
Context sets specify for a method \methodname the following: the methods in \abs{Succeeds} \emph{must} have run and \emph{must establish} the precondition as their postcondition. The methods in \abs{Overlaps} \emph{may} have run and \emph{must preserve} the precondition.

Additionally, \crowbar supports a variant of local session types, which are annotated with \abs{Local}. A local session type for \ABS is represented as a string and specifies calls, synchronization, sequence, repetition and alternative. E.g., the type from Ex.~\ref{ex:session} is specified with \abs{[Spec: Local("f!m.g!n.Put(result == 0)")]}.

Alternative is specified using \abs{+}, repetition with \abs{*}, suspension with $\xabs{Susp}(\phi)$ (where $\phi$ specifies the state before suspension, and analogously for calls \abs{!}) and synchronization with \abs{Get(e)}, where \abs{e} is the targeted expression.
Session types specify how \emph{roles} in a protocol communicate. The mapping of roles to fields is specified with \mbox{\abs{[Spec: Role("name",this.field)]}}.
For details we refer to~\cite{Kamburjan19,thesisek,KamburjanC18}.

The systems are independent: it is not necessary to specify a local type.

\paragraph{Nullability.}
\ABS implements a nullability type system: fields and variables may be marked as nullable \abs{[Nullable]} (default) or non-nullable \mbox{\abs{[NonNull]}}.
Additionally, nullability type inference can infer nullability for \emph{expressions} to statically detect whether a specific field or variable access always accesses a non-\abs{null} value.

\paragraph{Functional Layer.}
\ABS allows to define side-effect-free functions to simplify modeling of computations. Function-passing is not allowed.
They are, thus, a special case of Java \texttt{strictly\_pure}~\cite{jml,keybook} methods and we can 
verify pre-/postconditions for such functions by transforming them into a simple imperative structure.
To prove that a function \mbox{\abs{T f(p) = e;}} fulfills its pre- and postconditions $\mathit{pre},\mathit{post}$, we generate the following proof obligation.
\[\mathit{pre} \Rightarrow \big[\xabs{result = e;}\Vdash^{\alpha_{\mathsf{pst}}}\mathit{post}\big]\]
A contract generates a precondition that enables us to apply it to resursive calls. %, as each function is handled as a logical function symbol.
This is \emph{not} circular: the function is unrolled, so function contracts are analogous to recursive method contracts, and the precondition must be proven to use the postcondition. 
\crowbar can establish partial correctness of recursive functions.
\begin{example}
The following function results in the proof obligation below.

\noindent\begin{abscode}
[Spec: Requires(n >= 0)][Spec: Ensures(result >= 0)]
def Int fac(Int n) = if(n <= 1) then 1 else n*fac(n-1);
\end{abscode}

\vspace{-5mm}
\begin{align*}
&\big(\forall~\xabs{Int}~x.~x \geq 0 \rightarrow \mathtt{fac}(x) \geq 0\big) \wedge \xabs{n} \geq 0\\
\rightarrow &\big[\xabs{result = if(n <= 1) then 1 else n*fac(n-1);}\Vdash^{\alpha_{\mathsf{pst}}} \xabs{result} \geq 0 \big]
\end{align*}
\end{example}

%The condition on the function symbol is added to all other proof obligations.

%% file: middleend.tex
%\crowbar uses \emph{guided} symbolic execution, an extension of symbolic execution that considers external analyses and the specification while building the symbolic execution tree.
%
%\paragraph{Symbolic Execution.}
%As discussed above, symbolic execution engines build a proof tree to abstract over all concrete runs of a program.
%After SE has finished, all leaves are modality free and can be closed unsing first-order reasoning.

%BSE considers external analyses the specification while building the symbolic execution tree. 
\crowbar implements 2 kinds of behavioral guides (or just \emph{guides})\footnote{They are not related to assertion guided SE~\cite{GuoKWYG15}, which is a \emph{dynamic} technique.}: guides from the type system to reduce branching and guides from the specification.

\paragraph{Nullability Guides.}
An access that is shown to be non-\abs{null} by the type system is marked as such in the intermediate representation of \crowbar. 
BSE then does not generate a side-branch to prove that no NullPointerException is thrown. 
BSE still generates such a side-branch if the type system can not infer null safety. This is the default case for all method calls.

\paragraph{Specification Guides.}
Specification guides influence the shape of the tree and what side-branchs are generated by linking statements and specification more closely together than it is possible with post-conditions. While specification guides are already present in the \BPL calculus for object invariants, they are most prominent in two places in the aforementioned session type system:
%BSE also influences the shape of the tree itself. \crowbar implements (simplified) Local Session Types for \ABS~\cite{KamburjanC18}, which specify the order of communication actions (calls, synchronizations, suspensions, etc.) in a method.  
(1) During symbolic execution with a local Session Type as specification, \crowbar checks the order of actions by matching the active type action against the active statement. 
If an action cannot be matched, symbolic execution stops. 
(2) Branchings are matched non-greedy: if syntactic matching cannot determine what branch is taken, then the decision is delayed and all branches are considered.
The contained logical conditions are only generated at the very end, once the branch is identified and syntactic matching succeeded.
E.g., for %the behavioral modality
\[\big[\xabs{f!m(-10); return 0} \Vdash^{\alpha_{\mathsf{ST}}}\big((\xabs{r!m(i > 0)}.\xabs{r!n}) + \xabs{r!m(i < 0)}\big).\xabs{Put}\big]\] 
the constraint on \abs{i} is generated not when matching 
the call on \abs{m}, but at the end. %once the taken branch is determined in the next step%%.

%This exemplifies \emph{design} of specification syntax:
%The same order expressed as a first-order formula does not allow one to use such syntactic guides~\cite{thesisek}.
%LTL, which is also applied to SE~\cite{BeckertB13} also requires to first build the full SE tree.

%By stopping symbolic execution, \crowbar implements fail-early in a purely logical framework.
%Instead of completely analyzing the method to discover that the first action is faulty,
%\crowbar reports the first error of this kind directly. %before checking the generated side-branchs and continuing analysis.

%\begin{example}
Consider again Fig.~\ref{fig:gsetree}:
Node $(1)$ is omitted because of the nullability guide, 
while the other dashed nodes are omitted because of the specification guide.
%: the method is specified to make two calls, but executes three. Thus, the proof may fail-early once the superflous action (i.e., the third call) is detected.
%\end{example}

\paragraph{Modularity.}
%\crowbar implements two \BPL-calculi: one for local session types and one for contracts and invariants.
To avoid restrictions by committing to a special domain-specific language for rules~\cite{MitschP20}, adding new calculi requires that these are implemented directly in \crowbar.
To do so, data structures for the new specification languages and for their set of rules must be provided.

A rule must inherit from the class \abs{Rule} and (1) provide an abstract pattern and (2) implement the rule application.
An abstract pattern is an instance of a symbolic node that contains abstract variables. 
Pattern matching in \crowbar considers everything that implements the interface \abs{AbstractVar} an abstract variable. 
Pattern matching and matching condition (a mapping from abstract variables to concrete data) generation is implemented generically using reflection and does not need to be extended
by newly extended calculi.
In particular, it ensures that an abstract variable of class \abs{C} is matched with an instance of a subclass of class \abs{C} (that does not implement \abs{AbstractVar}).
The rule application itself is a method that takes a concrete symbolic node and matching conditions and returns a list of nodes.

%% file: backend.tex
The result of (B)SE is a symbolic execution tree. 
If one of the leaves is a symbolic node, then some behavioral guide aborted symbolic execution and verification fails.
Otherwise, all leaves are either \emph{logic nodes} or \emph{static nodes}.
\paragraph{Logic Nodes.}
Logic nodes are interfaces that contain no modalities and represent one proof goal.
\crowbar delegates these proof goals to SMT solvers by passing them the logic nodes translated into SMT-LIB.
If all goals can be closed, verification succeeds (up to possible static nodes), otherwise, the verification fails.

The translation has to take care of some differences between \BPL without modalities and SMT-LIB:
\begin{description}
\item[Heap] \BPL has a polymorphic heap, i.e., a heap maps fields to elements of some data types. 
SMT-LIB does not support polymorphic heaps: a heap can map fields only to elements of one data type.
To amend this, the translation splits the polymorphic heap into one monomorphic heap for each data type. 
Unused heaps are removed in a post-processing step.
\item[Partial Functions] 
\ABS allows partial pattern matching expressions, which throw an error if no branch matches.
\crowbar only considers partial correctness (and some exceptions), but must encode these expressions as 
total functions for SMT-LIB. This is done by a final `else' branch that returns an underspecified function symbol of fitting type.
This approach to handle partial functions when proving partial correctness is inspired by the \Key system.
\item[Updates] \BPL supports \emph{updates} to accumulate substitutions. These substitutions are applied before the formula is translated.
\end{description}

The translation to SMT-LIB is implemented in its own module -- implementing another backend does not require any changes to the symbolic execution.

\paragraph{Static Nodes.}
Static nodes are generic interfaces to communicate with external analyses. 
If a static node is present and all logic nodes can be closed, then the program is considered verified up to some external analysis on specific output.
For example, the context sets are output together with their heap precondition. 
The consistency of cooperative contracts~\cite{sao} and the adherence to the specified method order is checked externally and is an inherently non-local analysis.
Similarly, each class with local types outputs a static node for compositionality.
%resolution contracts output static node that contain a pair $(i,M)$, denote that the \abs{get} statement at position $i$ may only
%read from methods $\methodname \in M$. This is, due to future-passing, an inherently non-local analysis, and must be handled externally, e.g., by a points-to analysis.\ektodo{Change example to context sets!}

\begin{figure}[bt]
\centering
\renewcommand{\figurename}{Listing}
\begin{abscode}
class CeFrame {
  Int beats = 21239;  String s = "object_3";
  Unit ce() {              // Snippet from: heartbeat
    String req = "fut_1";  // Fut req = this.s!httpRequest();
    Int status = 5;        // Int status = (req).get;
    if((status == 200)){}
    else {// this.handleError(); Assume following assignments while blocked:
      this.beats = 21238;
    }
    println(toString(this.beats)); // Return stmt, evaluates to: 21238
    // Failed postcondition: (heap.beats>=old.beats) /\ (result=heap.beats)
    // Failed to show the following sub-obligations:
    // (select(anon(heap), this.beats)>=heap.beats)
  }
}
\end{abscode}
\vspace{-4mm}
\caption{The counterexample generated by \crowbar for the program in Lst.~\ref{fig:example}. }
\label{code:cegOutput}
\end{figure}

\paragraph{Counterexamples.}
\crowbar can generate a \emph{counterexample program} (CE) from each open branch in the SE tree.
Such a CE is a \emph{compilable} excerpt of the program under verification that is modified to illustrate the cause of the verification failure.
CEs allow users to quickly identify specification or implementation issues, without being exposed to the inner workings of \crowbar.

CEs are generated from the SE tree by traversing a single, open branch: based on the applied rule,
%(and additional information depending on the statement), 
each node in the tree can be transformed into a statement. 

The result of this reconstruction is an \ABS method which contains only statements relevant to runs represented by the open branch. 
However, not all structure of the original method is removed: if only one branch of an \abs{if} statement is relevant, the content of the other branch is not present in the CE, but
the \abs{if} statement itself is -- this provides the user more guidance to connect the input program with the CE.
Failure caused by loop invariants removes all information outside the loop (failure inside the loop) or inside a loop (failure outside of it).

To remove symbolic values and provide concrete values for variables and fields in the counterexample, \crowbar uses the model generated by \texttt{(get-model)} in the SMT solver from the failed logic node.
Similarly, this information is used to make the CE a stand-alone executable: 
any statements that require \emph{external} context (method calls, \abs{get}, \abs{await}, and object creations) are replaced with assignments that model the relevant \emph{internal} effects of the original statement.

With the knowledge extracted from the logic model and the annotations in the symbolic execution tree, \crowbar can further enrich the counterexamples with human-readable annotations. While these annotations are not strictly necessary for a working counterexample, they can aid the user in understanding the generated code. Among these additional annotations are detailed, static evaluations of return expressions, known-true statements at different points in the program, as well as a breakdown of the current proof obligations.

As an example, the CE generated by \crowbar for the input in Lst.~\ref{fig:example} is shown (slightly prettified) in Lst.~\ref{code:cegOutput}. 
Note the absence of the \abs{if} branch \emph{content} that is not on the problematic execution path in the counterexample as well as the annotations for method calls, return expressions and proof obligations. 
Concrete values are chosen by the SMT-solver.

%% file: conclusion.tex
\crowbar is used in the biggest verification case study for Active Objects:
A model extracted from C code~\cite{crowpaper} with 260 lines of \ABS code, with 5 classes (with invariants and creation conditions) and 5 interfaces (with 19 method contracts).
It also contains one function with a contract. 
The verification succeeds fully automatically.
In contrast, the previously biggest case study with \Key-\ABS~\cite{DinTHJ15} has 140 lines of code for 1 class and requires manual interaction.
The functional layer cannot be verified with \Key-\ABS and counterexample generation is not available (and restricted to pretty-printing models of single nodes in \Key-Java~\cite{keybook}).
Attempts to integrate \Key-\ABS with external tools~\cite{avocs} or new specification languages\footnote{\Key-\ABS only supports invariants written in ABSDL and ignores the functional layer.} have shown that its structure is not suitable for such tasks.
A discussion on the experiences made with \Key-\ABS is given in~\cite[Ch~2.3]{thesisek}.

\paragraph{Conclusion.}
We present \crowbar, a verification tool for Active Objects based on \emph{behavioral} symbolic execution.
By using behavioral guides, \crowbar integrates verification with static analyses and specification languages to influence the shape of the symbolic execution tree,
which opens the possibility for novel techniques in verification.
Our counterexamples are generated in the input language \ABS and presented to non-expert users without exposing the verification mechanism. 
\crowbar is suited for prototyping and implementing new specification approaches to distributed systems: it provides interfaces to implement new calculi and communicate with external analyses. \crowbar has been used for the biggest case study using Active Objects and is the first implementation of BSE.

\paragraph{Future Work.}
We plan to 
%implement symbolic traces~\cite{DinHJPT17}, 
%empirically compare different approaches to specification
%and to 
pass static nodes to dynamic monitors, combining dynamic and static verification~\cite{ChimentoAPS15,LSGD}.
Furthermore, we explore the possibility to export the the proof obligations for the functional layer, which are not specific to \BPL, to specialized provers, e.g., Why3~\cite{FilliatreP13}.
%, where the static nodes are input to dynamic monitors.
%, and with the verification of Hybrid \ABS~\cite{hscc}, based on \emph{differential} dynamic logic~\cite{Platzer12b}.